# Lithospheric delamination beneath the southern Puna plateau resolved by local earthquake tomography

**Jing Chen[1,2], Sofia-Katerina Kufner[2,3], Xiaohui Yuan[2], Benjamin Heit[2], Hao Wu[1], Dinghui Yang[1], Bernd Schurr[2], and Suzanne Kay[4]**

[1]Tsinghua University, Beijing, China.

[2]Deutsches GeoForschungsZentrum GFZ, Potsdam, Germany.

[3]British Antarctica Survey, Cambridge, UK.

[4]Cornell University, Ithaca, USA

Corresponding author: Jing Chen (jing-che16@mails.tsinghua.edu.cn)

**Key Points:**

- A local earthquake tomography is performed in the southern Puna plateau and adjacent region
- A high mantle Vp anomaly beneath Cerro Galan provides evidence for lithospheric delamination
- A low Vp zone is observed beneath Ojos del Salado, reflecting the melt supplied by slab dehydration reaction at two depths

## Abstract

We present a local earthquake tomography to illuminate the crustal and uppermost mantle structure beneath the southern Puna plateau and to test the delamination hypothesis. Vp and Vp/Vs ratios were obtained using travel time variations recorded by 75 temporary seismic stations between 2007 and 2009. In the upper crust, prominent low Vp anomalies are found beneath the main volcanic centers, indicating the presence of magma and melt beneath the southern Puna plateau. In the lowlands to the southeast of the Puna plateau, below the Sierras Pampeanas, a high Vp body is observed in the crust. Beneath the Moho at around 90 km depth, a strong high Vp anomaly is detected just west of the giant backarc Cerro Galan Ignimbrite caldera with the robustness of this feature being confirmed by multiple synthetic tests. This high velocity body can be interpreted as a delaminated block of lower crust and uppermost mantle lithosphere under the southern Puna plateau. The low velocities in the crust can be interpreted as having been induced by the delamination event that triggered the rise of fluids and melts into the crust and induced the high topography in this part of the plateau. The tomography also reveals low velocity anomalies that link arc magmatism at the Ojos del Salado volcanic center with slab seismicity clusters at depths of about 100 and 150 km and support fluid fluxing in the mantle wedge due to dehydration reaction within the subducted slab.

## 1. Introduction

The southern Puna plateau constitutes the southern termination of the Central Andean Altiplano-Puna plateau (Figure 1a), which after Tibet is the world's second largest continental plateau and differs from Tibet in having been formed on an active continental subduction margin, specifically by the subduction of the oceanic Nazca plate beneath the continental South American plate. The major geophysical, geologic and magmatic features and the evolution of the southern Puna have been summarized by a number of authors (e.g., Isacks, 1988; Allmendinger et al., 1997; Oncken et al., 2003; Kay & Coira, 2009). When compared to the northern Puna and Altiplano plateau, the southern Puna plateau is characterized by (a) a thinner crust and thinner mantle lithosphere (e.g., Whitman et al., 1996; Heit et al., 2007, 2014); (b) being in a region where the subducting slab is changing from a steeper dip in the north to a shallower dip in the south and where a seismicity gap exists in the subducting slab (e.g., Cahill & Isacks, 1992; Mulcahy et al., 2014); (c) having a distinctive sedimentary, magmatic, and structural history (e.g., Kley & Monaldi, 1998, Kay et al., 1999 and references therein). These characteristics make the southern Puna plateau an excellent natural laboratory to study the processes that form a high plateau within the frame of an active continental margin subduction zone.

The northern and southern Puna are separated by the northwest trending Olacapato-Toro-Lineament (O-T-L) near 24.5°S. To the east, the boundary of the southern Puna is structurally limited by the Santa Barbara system north of 27°S and by the Sierras Pampeanas to the south. In detail, the Santa Barbara system is characterized by predominantly west verging, relatively high-angle thrust faults that are largely inverted Cretaceous normal faults and the Sierras Pampeanas are thick-skinned basement uplifts that are bounded by high angle reverse faults (Allmendinger & Jordan, 1997; Kley & Monaldi, 1998, 2002). To the west, the southern Puna is bounded by the Andean Neogene Central Volcanic Zone (CVZ). Within the region of the southern Puna are a series of NW-trending zones of lithospheric weakness along which the largest volcanic centers are located (Figure 1b, Archibarca, Culampaja, and Ojos del Salado lineaments) (e.g., Heit et al., 2014).

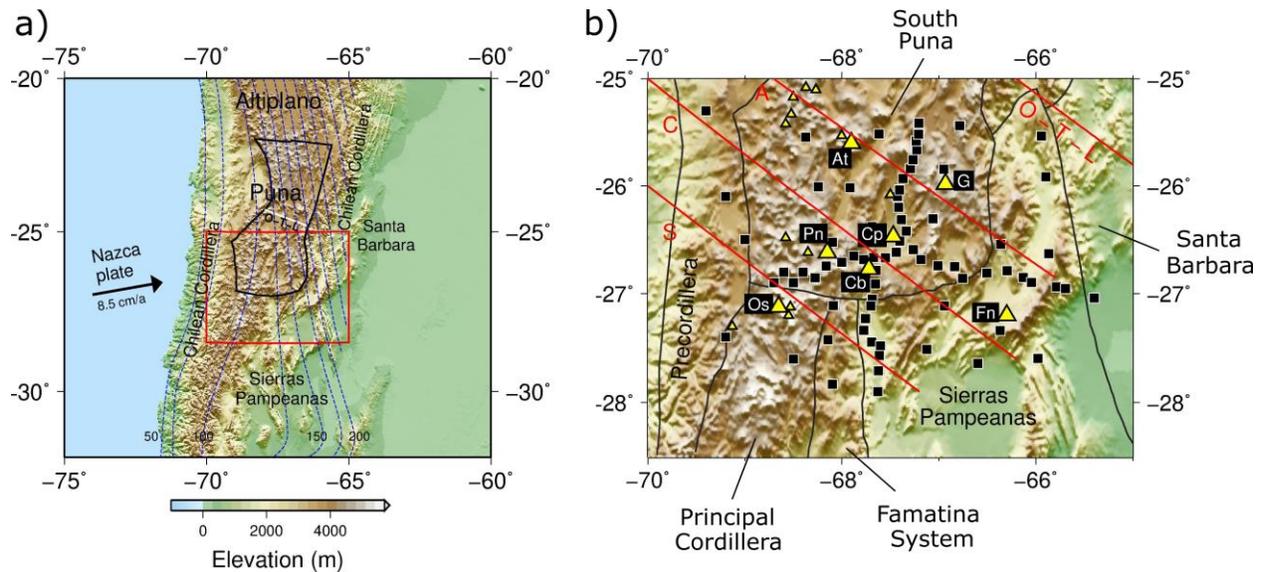

**Figure 1.** (a) Map of main tectonic features in the Puna plateau and adjacent regions from 20°S to 32°S. Blue dashed lines are the Wadati-Benioff contours (Cahill and Isacks, 1992). The black line outlines the Puna plateau, which is separated to the northern and southern parts by the O-T-L (Bianchi et al., 2013). The Nazca plate convergence rate is from the integrated NUVEL-1 model (DeMets et al., 1990). The red rectangle marks the working area. (b) Distribution of stations (black squares) and active volcanoes (yellow triangles) in the southern Puna plateau. Black solid lines denote the tectonic boundaries. Major volcanic centers are labeled (At: Antofalla; G: Cerro Galan Caldera; Cp: Carachi Pampa; Pn: Cerro Peinado; Cb: Cerro Blanco Caldera; Os: Ojos del Salado; and Fn: Farallon Negro). Main lineaments are represented by red lines (A: Archibarca lineament; C: Culampaja lineament; S: Ojos del Salado lineament; O-T-L: Olacapato-Toro-Lineament).

The Puna plateau has an average altitude of 4.2 km, which is 1 km higher than the Altiplano plateau (see Whitman et al. 1996). Receiver function studies show that the crust beneath the southern Puna plateau has a thickness of 50-55 km and is on average 10 km thinner than the northern Puna (~60 km) and 20 km thinner than the Altiplano (~70 km) (Beck et al., 1996; Yuan et al., 2002; Wölbern et al., 2009; Heit et al., 2014). Additionally, the crust to the eastern border of the Puna plateau is slightly thicker than in the central part of the plateau, which could suggest that the crustal shortening is concentrated at the border of the plateau and becomes progressively thinner in the Eastern Cordillera, the Santa Barbara System, and the Pampean Ranges (Kley & Monaldi, 1998; Kley et al., 1999).

Young mafic volcanism distinguishes the southern from the northern Puna plateau (Kay et al., 1994). Especially, the large ignimbrite deposits in the Cerro Galan caldera indicate large quantities of crustal melts in the southern Puna plateau. Based on the geochemical and petrological fingerprint of the volcanic rocks, Kay and Kay (1993) and Kay et al. (1994) proposed a model of delamination of thickened crust and lithosphere model under the southern Puna plateau. Such a model is capable to explain the presence, distribution and chemistry of the mafic volcanic rocks, the change from a thrust-dominated to a mixed thrust, reverse and normal fault regime (Marrett et al., 1994), the large volumes of ignimbrite deposits and the high topography over a relatively thin crust. Specifically, the Cerro Galan caldera ignimbrites were suggested to result from melting caused by mafic melts that were generated in association with lithospheric delamination being injected into the crust. The delamination and crustal melting may be related to the steepening of a formerly shallower subducted slab (Kay & Coira, 2009; Kay et al., 2010).

Based on local tomography results to the immediate north of our study region, Schurr et al. (2003 and 2006) described ascending paths for the fluids and melts at 24°S in the northern Puna plateau. As did Coira and Kay (1993) on the basis of the chemistry of the magmatic rocks, they suggested that these melts were induced by thermal instability of the lower crust and uppermost mantle and concluded that piecemeal delamination could have been responsible for these anomalies.

Since these pioneer studies, many geophysical studies have focused on the southern Puna and tested the delamination hypothesis. A high velocity block in the upper mantle beneath the Cerro Galan caldera was detected by teleseismic body and surface wave tomography (Bianchi et al., 2013; Calixto et al., 2013) and by attenuation tomography (Liang et al., 2014). A high Vp/Vs ratio and a low shear wave velocity anomaly was observed in the crust beneath Cerro Galan by using receiver functions and ambient noise correlation, which indicated partial melt or the presence of a magma chamber consistent with the delamination hypothesis (Heit et al., 2014; Ward et al., 2017; Delph et al., 2017). This is consistent with a pronounced high attenuation zone in the crust beneath Cerro Galan (Liang et al., 2014). Calixto et al. (2013) showed evidence for the delaminated block by surface wave tomography and suggested it is located beneath the Moho and above the slab in the region where there is a gap in seismicity at the depth of the subducted slab (Mulcahy et al., 2014). Although all these studies agree on the existence for a delaminated block, they fail to provide conclusive evidence of the delamination process and the accurate location of the delaminated block. This is mainly due to a lack of resolution of these studies at crustal and shallow upper mantle levels as they all use teleseismic waves.

In this study, we perform a local earthquake tomography and aim to obtain high resolution images of the crust and the uppermost mantle by making use of travel time data recorded by the 75 seismic stations in the southern Puna seismic network that operated between 2007 and 2009. P-wave velocities and Vp/Vs ratios in the crust and uppermost mantle are obtained and analyzed to test the delamination hypothesis as a working model involving the subduction of an oceanic plate and the evolution of the Andes as a result of orogenic processes.

## 2. Data and Methodology

### 2.1. Seismic network and input data

A passive source seismic array consisting of 75 stations was deployed across the southern Puna plateau (Heit et al., 2007; Sandvol & Brown, 2007) and operated between 2007 and 2009. The network was arranged into two orthogonal profiles along north-south and west-east directions with small inter-station spacing (~10 km) and additional stations with spacing between 35 km and 50 km in between (Figure 1b).

We use a groomed version of the local earthquake catalog from Mulcahy et al. (2014) to calculate a local earthquake tomography. The catalog by Mulcahy et al. (2014) contains 1903 earthquakes (25077 P- and 14059 S-picks) located in a regional 1D velocity model. In order to eliminate the influence of unreliable events and observations, strict selection criteria are applied. We only include events that are recorded by more than 6 stations, whose azimuthal gap is smaller than 315° and whose initial root mean square (RMS) travel time residual prior to relocation is smaller than 1.5 s. Additionally, low-quality picks whose residuals are larger than 3 s for P-wave or 5 s for S-wave are excluded from the analysis. The final inversion dataset consists of 1383 earthquakes recorded at 75 stations and constrained from 21781 P-wave and 12441 S-wave picks (Figure 2a), yielding a good ray coverage within a domain ranging from 64 to 71ºW in longitude and from 24 to 29ºS in latitude, down to a depth between 100-200 km depth (Figures 2b and 2c).

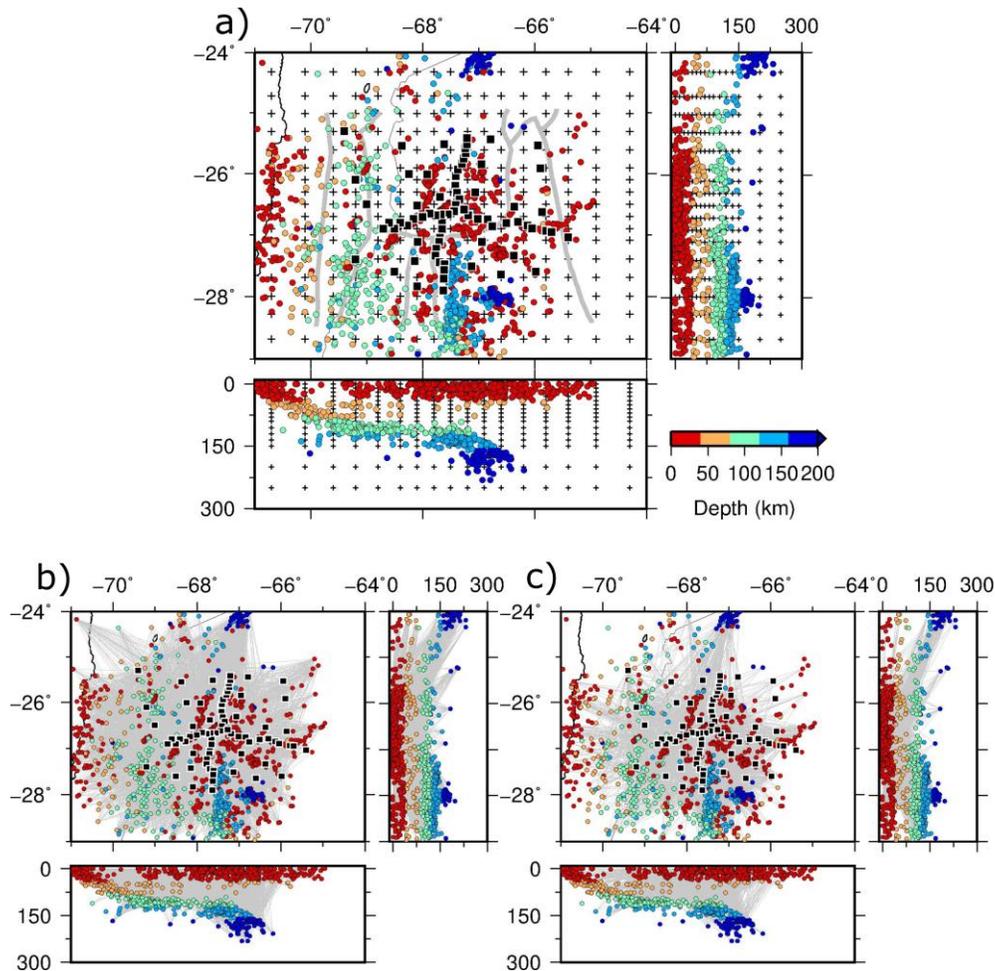

**Figure 2.** (a) Map of stations (squares) and earthquakes (circles). Earthquakes are color-coded by the depth. Black crosses represent the grid nodes for the 3D inversion. Gray lines are the same tectonic boundaries from Figure 1. The black line denotes the coastline. (b) and (c) P-wave and S-wave ray paths (gray lines).

## 2.2. 1D starting velocity model

A starting model of Vp and Vp/Vs ratios is needed for the local earthquake tomography. Generally, this nonlinear inverse problem is approximately formulated as a linear function to be solved (Kissling et al., 1994). Thus, 3D inversion strongly depends on the starting model. In order to obtain a suitable starting model for the 3D local earthquake tomography, a minimum 1D Vp model is derived based on a groomed high-quality dataset of the event catalog (Figure 3a; 247 events, 6511 P-picks) by using the software VELEST (Kissling et al., 1995). Earthquakes in this high-quality dataset are recorded by more than 15 stations so that they can be relocated robustly, which can make sure of the stability of 1D inversion.

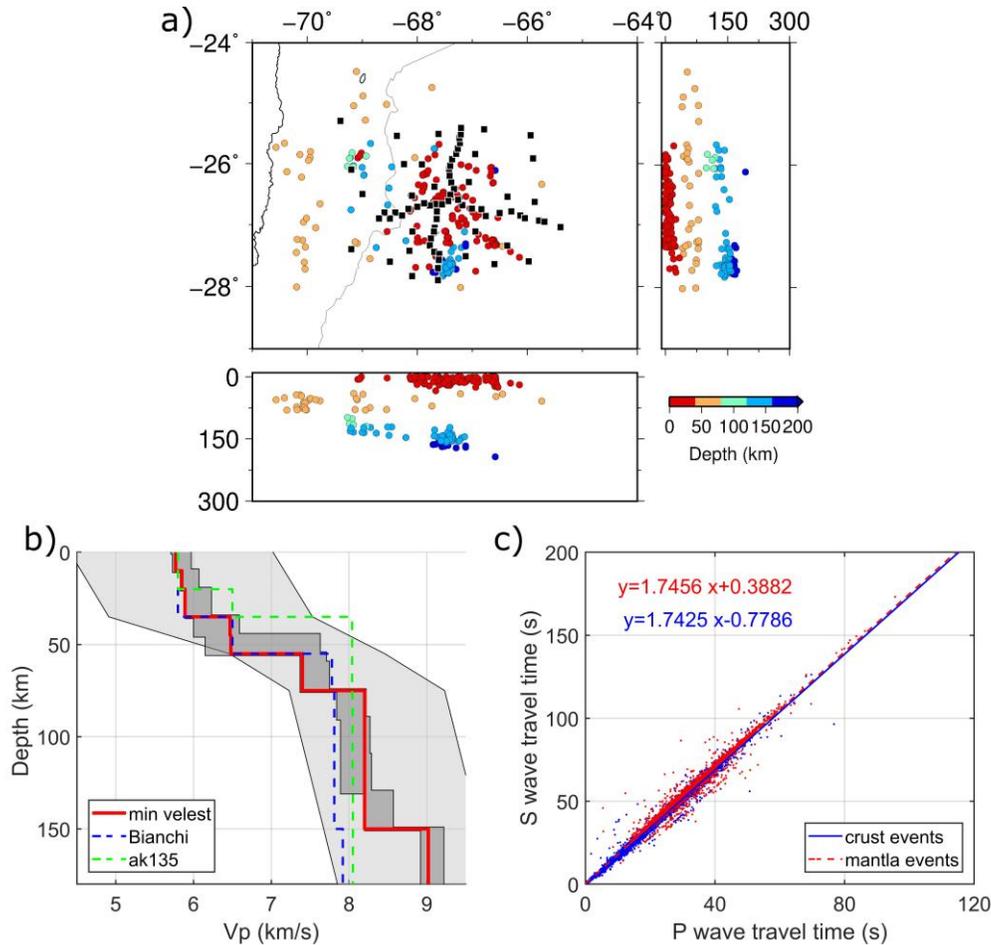

**Figure 3.** (a) Earthquakes (circles) and stations (black squares) involved in the inversion for the 1D Vp model. (b) The minimum 1D Vp model (red line) obtained by the 1D inversion. Blue and green dashed lines denote models of Bianchi et al. (2013) and the AK135 model, respectively. The light gray shaded area represents the range of input models tested during the inversion for the minimum 1D model. The dark gray domain represents the range of probable output models after inversion. (c) Wadati-diagram based on the earthquake catalog. A constant Vp/Vs ratio of 1.74 is derived by linear regression with earthquakes in the crust and in the mantle, respectively.

During the 1D inversion, we allow for the simultaneous inversion of station corrections, event locations, and P-wave velocity model. We firstly test a wide range of input 1D models (Figure 3b, light gray domain) with different Moho depths and P-wave velocities in different sampling layers. Then, those layers which have little contrast to their surroundings are subsequently combined, leading to models including 11 layers from 0 km to 200 km. Finally, the minimum 1D Vp model is determined with the smallest RMS residual (Figure 3b, red line). As most output models collapse near the finally chosen minimum 1D model (Figure 3b, dark gray domain), we are confident that it represents a robust global minimum. The Vp/Vs ratio is constrained from the Wadati-diagram and set to 1.74 (Figure 3c). After the relocation with the minimum 1D Vp model, the RMS residual of dataset decreases to 0.340 s (Figure S1a), which is much smaller than the RMS residual (0.485 s) of dataset relocated by using Bianchi's model (Bianchi et al., 2013, Mulcahy et al., 2014). The station corrections show relatively large variations (Figure S1b and S1C; maximal/minimal values of 0.90/-1.55 s for P-picks and 1.09/-2.22 s for S-picks, respectively), correlating with main tectonic features (e.g., similar sign and magnitude in the southern Puna (in the central part) and the Sierras Pampeanas in the southeast), hinting towards different subsurface velocities within these units which will be resolved in the 3D local earthquake tomography.

## 2.3. Local earthquake tomography

The local earthquake tomographic inversion is performed with the software SIMULPS (Thurber, 1983; Evans et al., 1994), in which Vp, Vp/Vs ratios, hypocenter coordinates and origin times are inverted from the residuals between observed and theoretical travel time using the least-squares method (Menke, 1984). To describe Vp and Vp/Vs ratios, we set $19\times19\times17$ nodes in a rectangular grid to represent our study region of $700\times550\times200$ km along the longitude, latitude, and depth dimensions with center at 26.5°S/67.5°W (Figure 2a, black crosses). Node spacing is chosen considering distribution and density of rays with around 20 km at the center of the region and increased to 60 km towards the edges. Nodes are set every 10 km in depth in the crust and uppermost mantle (<90 km). The separation increases to 15 km for the deeper region (90–150 km). The initial Vp and Vp/Vs ratios are set on these fixed nodes according to the minimum 1D model in Figure 3b (red line), forming the initial input 3D model. In order to weaken the instability and non-uniqueness of solutions, suitable damping parameters with respect to Vp and Vp/Vs ratios are determined from trade-off curves in a multi-step procedure: First we perform the inversion with a series of Vp damping and over-damped Vp/Vs; then a preferred Vp damping parameter is determined from the trade-off curve, striking a balance between the model and data variance (Figure S2a). The same procedure is performed for Vp/Vs damping with fixed preferred Vp damping (Figure S2b).

Ten iterations were performed during the whole inversion. In the first two iterations, the initial earthquake hypocenters were fixed to ensure the stability of the inversion. Both P- and S-variance decreases strongly in the first two steps (84% for P-variance and 87% for S-variance respect to the total reduction in Figure S3). The variance still decreases in the next 8 iteration steps and finally becomes steady. In addition to the input and procedure described above, we ran similar inversions with slightly different grid spacing and event catalogs of fewer higher quality events (i.e., azimuthal gap < 240°). Though some anomalies at the edge or below the crust are not recovered due to fewer ray coverage, the revealed anomalies by using higher quality events are consistent with the origin inversion results (Figure S4). This comparison indicates that the anomalies discussed below are not artefact of grid setting or input velocity model.

## 3. Results

### 3.1 Inversion of real data

Our inversion resulted in a 3D P-wave velocity model in the crust and the uppermost mantle above the Nazca subducting slab in the southern Puna plateau and adjacent areas, which are presented as horizontal sections (Figure 4) and vertical profiles (Figure 5). Regions of good resolution defined by the spread value (e.g., Toomey & Foulger, 1989) are determined from the analysis of synthetic tests (see details in the next section).

## a) Crust

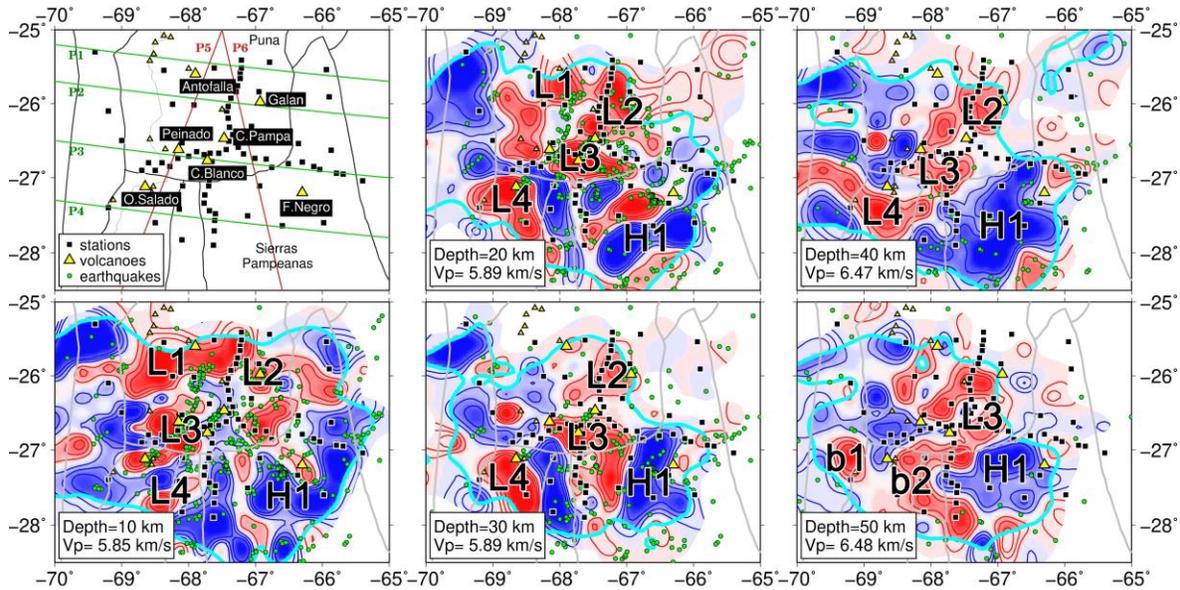

## b) Mantle

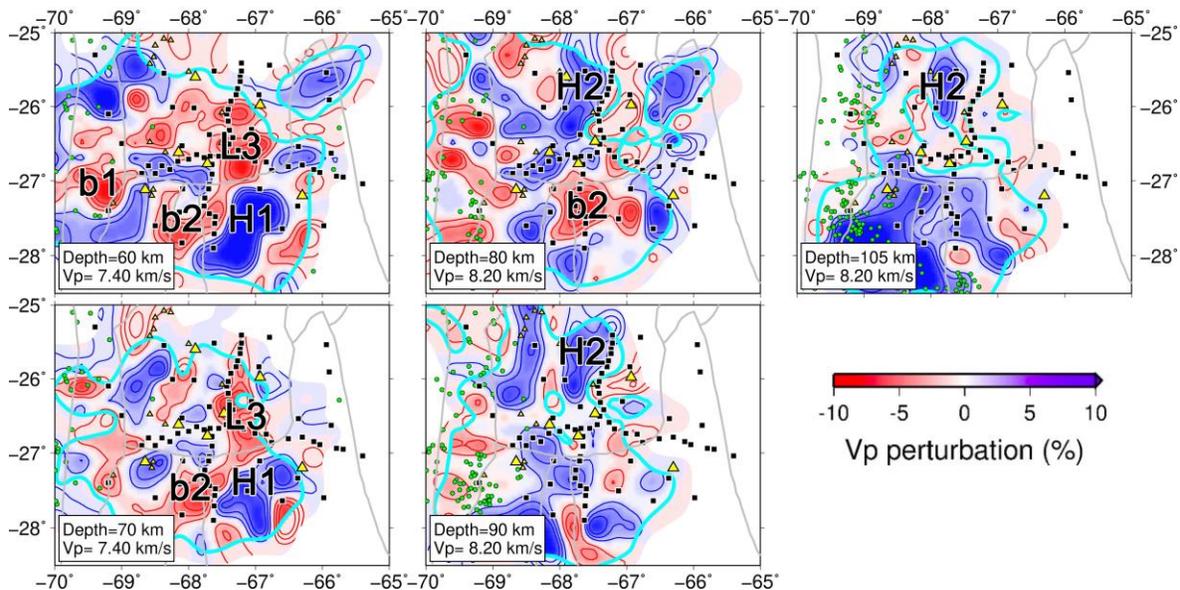

**Figure 4.** Horizontal sections of the inverted Vp perturbation model at different depths. Depth and reference Vp are indicated in the lower left box in each section. The well-resolved region is defined by the light blue lines, which are derived by spread value contours from the synthetic tests (Figures 6 and 7). Stations (black square) and main volcanic centers (yellow triangles) are plotted in each section. Green circles mark earthquakes within 20 km around the depth of each section. Gray lines represent tectonic features as shown in Figure 1. Main velocity anomalies are labeled for discussion. Green and red line in the upper left figure mark locations of the vertical sections show in Figure 5.

## a) W-E profiles (P1-P4)

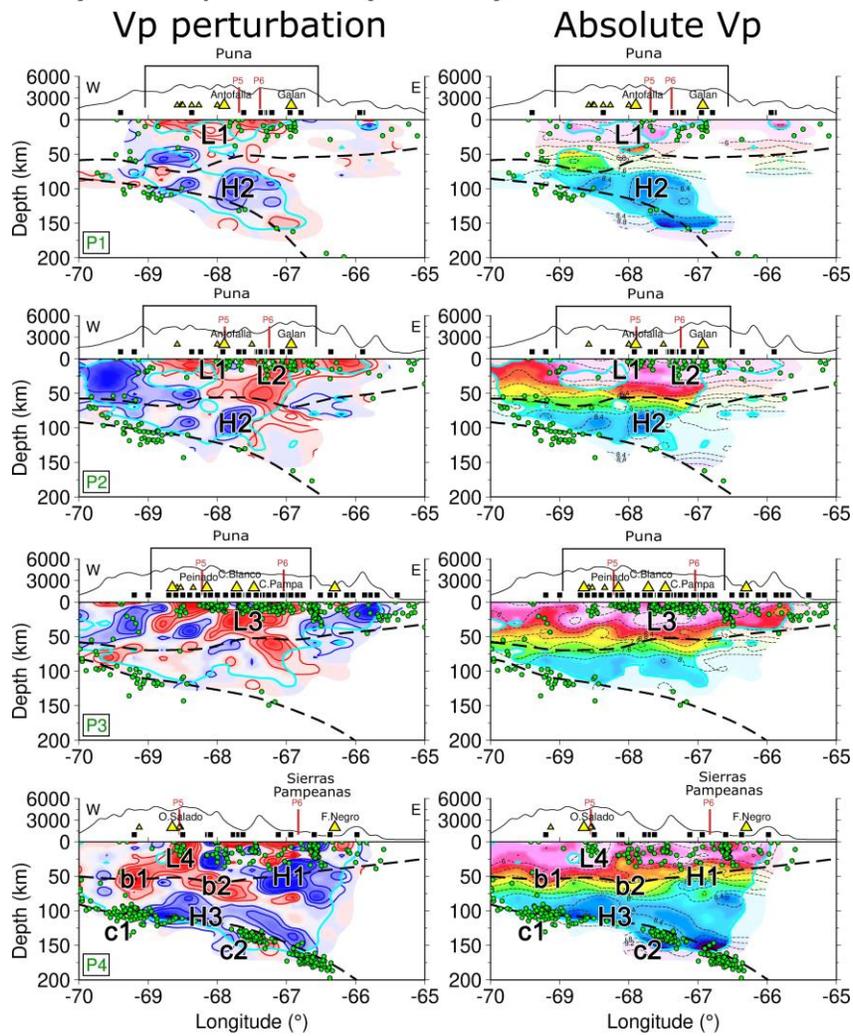

## a) N-S profiles (P5-P6)

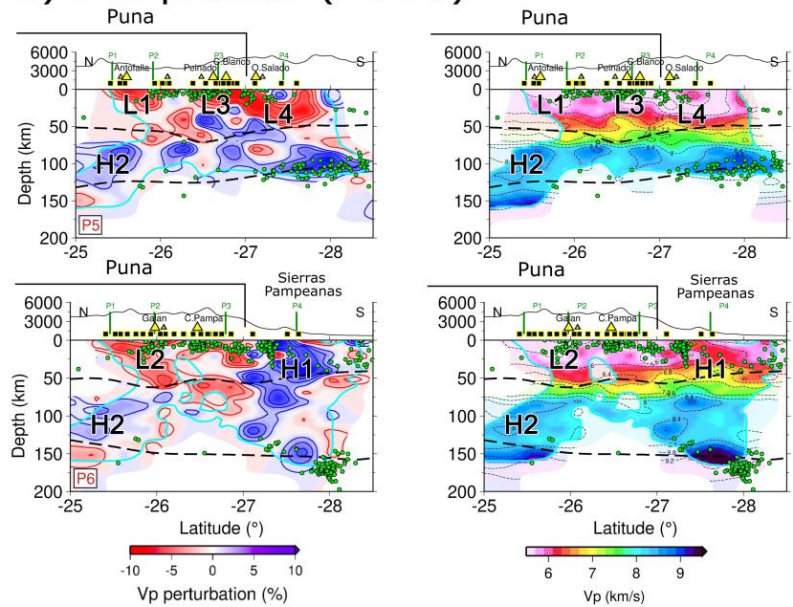

**Figure 5.** Vertical profiles of the perturbation and absolute Vp from the inverted model (locations are indicated in Figure 4a). Dashed lines mark the Moho from Heit et al. (2014) and the subducted slab from Mulcahy et al. (2014). Earthquakes (green circles), stations (black squares) and volcanoes (yellow triangles) are within 50 km along the profile. All other features are the same as in Figure 4.

In the upper crust above 20 km depth (Figure 4a), the southern Puna plateau is represented by a low Vp anomaly that matches with tectonically defined boundary of the plateau. By contrast, a high Vp zone "H1" is located to the southeast, below the Sierras Pampeanas block. Below the Puna Plateau, several low Vp zones are located just beneath main volcanic centers (e.g., "L1" beneath Antofalla; "L2" beneath Cerro Galan; "L3" beneath Cerro Peinado, Cerro Blanco and Carachi Pampa; "L4" beneath Ojos del Salado; Figures 4a and 5). The low Vp zones beneath Cerro Peinado, Cerro Blanco, and Cerro Pampa additionally extend to the east from crust to the mantle (Figure 5a, P3). Two further low Vp branches connect to the low Vp zone beneath the Ojos del Salado volcano (Figure 5a, P4): One extends to the west and stays at 70 km depth ("b1"), which might be associated with the earthquake cluster in the slab at 100 km depth ("c1"). The other anomaly extends east down to the uppermost mantle along the subducting slab ("b2"), towards the earthquake cluster at 150 km depth in the slab ("c2").

At upper mantle depths, a strong high Vp zone "H2" is detected from 80 km to 130 km within the well-resolved region beneath the Puna. "H2" is a large high Vp anomaly ranging from 25 to 26.3ºS, sitting above the subducting slab and beneath Moho (see Figure 4b at 80, 90 and 105 km depth). "H2" is also clearly visible along profiles P1, P2 (Figure 5a) and P5 (Figure 5b). The uppermost limit of the subducting slab is indicated by high Vp zone "H3" in the southernmost part of the study region (Figure 5, P4), where it can be recovered due to the nearby earthquake clusters ("c1", "c2").

### 3.2 Synthetic tests

Two groups of synthetic tests were performed to analyze the reliability of the features described in section 3.1. The first test was a standard checkerboard test to analyze the general resolution with respect to the geometry of stations and earthquakes in the dataset. Based on this test, we use the spread value to define well resolved model domains. The spread value is calculated from the model resolution matrix, measuring the smearing of model parameters (Toomey et al., 1989). A suitable threshold for the spread value is determined from the comparison between the well-resolved region and different spread value contours. Another set of synthetic tests was performed with a more complex synthetic input model geometry that resembles the inversion results obtained from the real data. It helps us to analyze the reliability and resolvability of the prominent features in the inversion results of the real data.

Both groups of synthetic tests are created according to the following procedure. After the creation of the input synthetic velocity models, theoretical travel times of all events in the dataset are calculated by forward ray tracing. Gaussian noise is added to these synthetic catalogs (standard deviation of 0.1 for P-picks and 0.2 for S-picks). These noised catalogs and a linear 1D model are then used to start the 3D inversion. During the inversion of the synthetic checkerboard test, suitable damping values need to be reassessed via trade-off curves as checkerboards do not represent the realistic distributions of anomalies, but the size and amplitude of the anomalies in the inversion result strongly depend on the choice of the damping. The procedure used is similar to the inversion of real data. For the inversion of the complex test, we used the same damping parameters as the real data for inversion for consistency.

The checkerboard is designed with staggered positive and negative velocity perturbations whose amplitude is ∓10% for Vp and ∓5% for Vp/Vs ratio (Figure 6). Each anomaly cube is similar in size (70×70×30 km), including two or three grid nodes in latitude, longitude and depth direction. A suitable spread value contour (light blue lines) is selected to describe the well-resolved domain, within which most anomalies are distinguishable.

## a) horizontal profiles

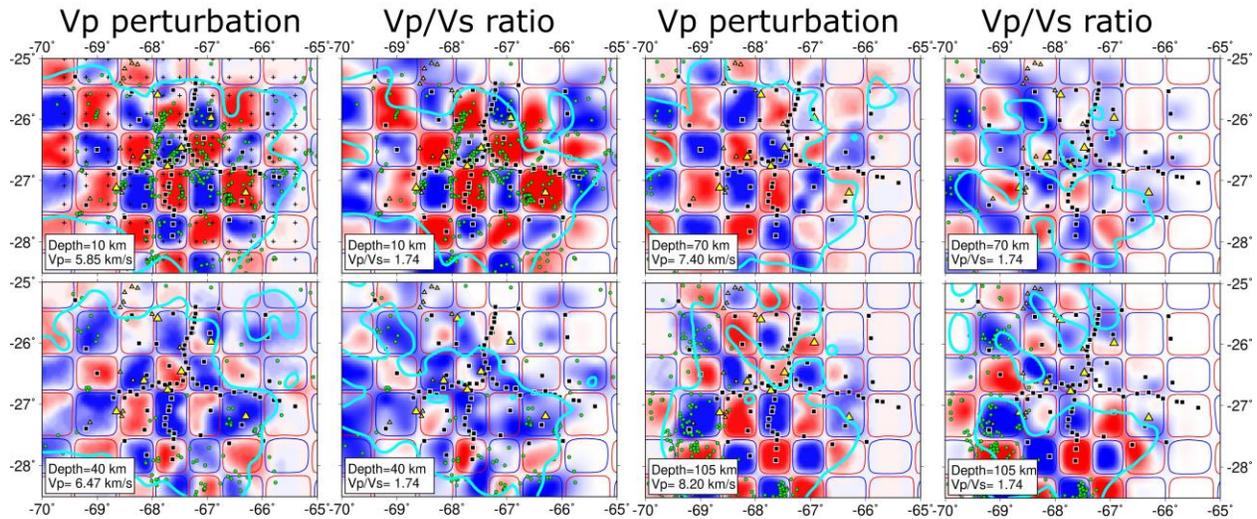

## b) vertical profiles

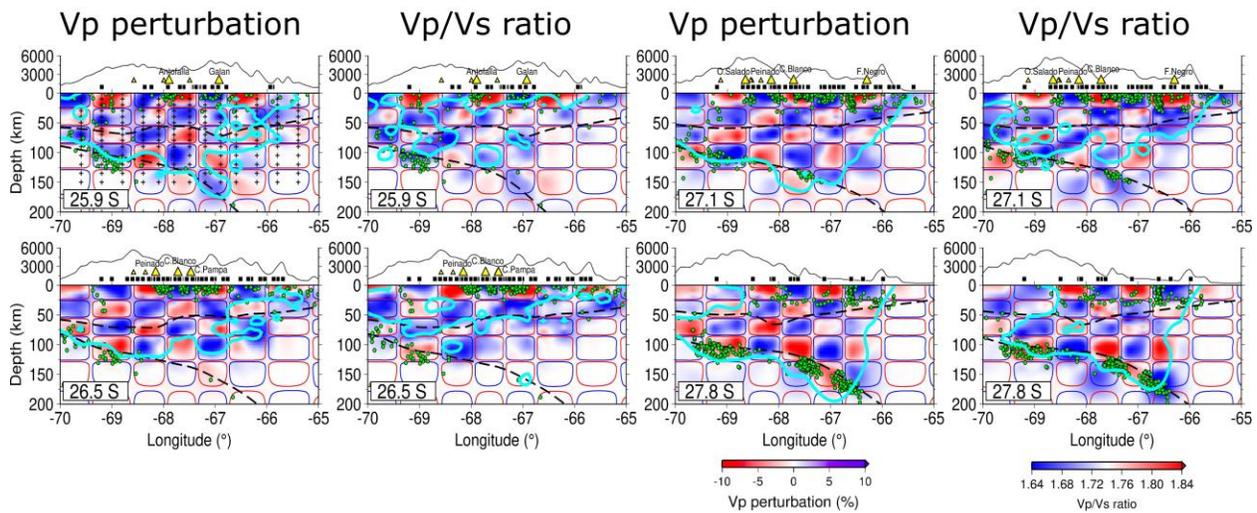

**Figure 6.** Checkerboard test for resolution of Vp and Vp/Vs ratio. Input positive and negative anomalies are indicated by blue and red lines, respectively. Light blue lines mark contours of spread value of 6 for Vp and 5.5 for Vp/Vs ratio, within which most anomalies are distinguishable. These contours are chosen to define the well-resolved domain for the real data inversion. (a) Horizontal sections. Depth and reference Vp and Vp/Vs ratio are indicated in the lower left box. (b) W-E vertical profiles along different latitudes, indicated in the lower left box.

For the second test, two velocity anomalies located within the model domain and identified from the real data inversion or from tectonic background knowledge are added in the synthetic model: a low Vp anomaly located in the southern Puna above 20 km depth, and a high Vp region "H1" in the southeast beneath the Sierras Pampeanas (Figure 7). Based on the slab geometry from Mulcahy et al. (2014), the subducted slab "H3" is introduced in the model. In addition, this synthetic model (Figure 7) also includes two features that appear as clear anomalies in the real data, but are rather small-scaled or at the edge of our model domain: the high Vp zone "H2" (e.g., Figure 5, P1 and P2), and the low Vp branches "L4" beneath Ojos del Salado (Figure 5, P4). For comparison, we performed another synthetic test (Figure S5) without these two anomalies. At the position of the removed features, no anomalies are recovered in the inversion results, which indicates that "H2" and "L4" are not a result of smearing but likely true anomalies.

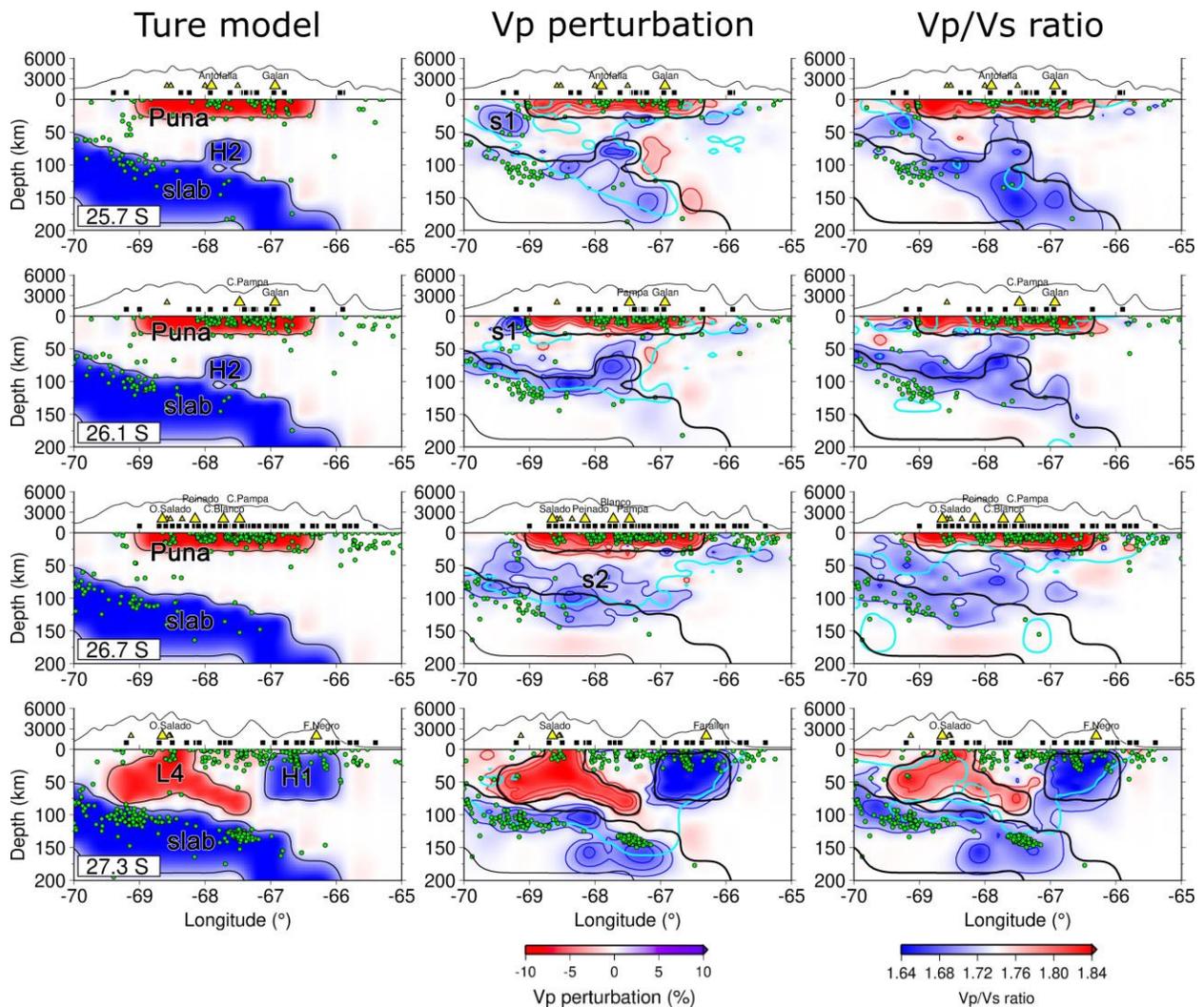

**Figure 7.** Anomaly resolution test along the 4 E-W vertical profiles. Positive and negative anomalies simulate the perturbations of Vp and Vp/Vs ratio obtained from the inverted model. Black lines denote the boundaries of input synthetic anomalies for reference. Light blue lines denote the contours of spread value of 6 for Vp and 5.5 for Vp/Vs ratio.

In the second synthetic test, the low Vp anomaly in the upper crust, representing the southern Puna plateau, can be well resolved. The strong Vp anomaly "H2" in the uppermost mantle above the subducting slab is distinguishable from the slab and the background mantle structure (profiles at 25.7°S and 26.1°S in Figure 7). At 27.3°S, thanks to large amounts of earthquakes in the subducting slab, the low Vp anomaly "L4" in the west and the high Vp zone "H1" in the east are recovered well. Additionally, the high Vp region along the earthquake clusters indicates the position of the subducting slab. However, although the main anomalies are well resolved, some regions, which we disregarded when introducing the results in Section 3, exhibit intense smearing. At 25.7°S and 26.1°S, an artificial high Vp anomaly ("s1") is introduced in the upper crust to the east of the low Vp Puna anomaly. The artifact is likely introduced by uneven distribution of earthquakes associated with distinct stations at the edge of our study region and a smaller number of events used from the slab seismicity gap. Almost all the stations are located east of 69°W. The region west of it, where anomaly "s1" is located, is only sampled by a small number of rays from nearly the same directions, which erroneously project the high velocities from the slab to the crust above it. Another highly smeared region ("s2") exists at 26.7°S above the slab where the high Vp slab structure is smeared to shallower depths due to poor ray coverage in this domain.

Based on the synthetic tests and inversion results, we can discriminate which features of the P-wave velocity model are robust to make interpretations. However, the tests also showed that 3D Vp/Vs ratio structures are not robust enough to be used for interpretation. This is probably due to fewer rays crisscrossing the model domain and higher picking uncertainties in S arrival times (Lange et al., 2018). The well resolved domain of Vp/Vs results are too small and disconnected to allow for interpretations. Thus, we will not discuss the Vp/Vs model any further but provide it in the appendix (Figures S6 and S7), because the it is still valuable for event relocation.

## 4. Discussion

### 4.1. Lithospheric delamination beneath the southern Puna plateau

Delamination is commonly referred to as the peeling away of the lower portion of the continental lithosphere and its sinking into the asthenosphere. Delamination is generally thought to be an ephemeral event which lasts only a few million years (Houseman et al., 1981; Schott & Schmeling, 1998) and most likely occurs when mantle lithosphere along with dense thickened crust becomes gravitationally unstable and detaches (Kay & Kay, 1993). When delamination occurs, a sequence of observations follows such as rapid uplift and an increase of magmatic activity, which are recognizable in the southern Puna plateau especially near the Cerro Galan caldera (Kay et al. 1994, 2011). The southern Puna is on the average higher than the northern Puna and has less crustal shortening ratio (Isacks 1988; Kley & Monaldi, 1998). Additionally, a giant eruption of the Cerro Galan ignimbrite complex occurred at 2.13 to 2.06 Ma (Kay et al. 2011). A similar case can be made for the northern Puna ignimbrites in accord with the seismic images and discussion in Schurr et al. (2006, see further interpretation in Kay & Coira, 2009).

Kay and Kay (1993) proposed a delamination model in the southern Puna plateau to explain the presence of mafic rocks, the large volume of ignimbrites, the thin crust and lithosphere, and the high topography with a large deficit in crustal shortening. In this model, the lower crust becomes thick and dense due to the compression that caused the Puna uplift. Accompanied with dense mafic residues created at the base of the crust, a critical amount of shortening makes the lithosphere denser than the underlying asthenosphere and gravitationally unstable, leading to the delamination event in the southern Puna plateau. As a result, the hot mantle produces mantle melts that rises into the crust and generate the hybrid mantle-crustal melts that erupt as ignimbrite from mid to upper crustal magma chambers (see Kay et al. 2010, 2011). This sequence of events might cause the uplift of the crust, the high topography, and the magma chamber beneath Galan.

Our high resolution tomographic images of the velocity distribution in the crust and uppermost mantle of the southern Puna plateau illuminate tectonic features anticipated in the above described delamination model. In the crust, a low velocity region within the southern Puna plateau is observed (Figure 4a, L1–L3), which correlates well with the "Southern Puna Magmatic Body" (SPMB) described by Bianchi et al. (2013). However, due to the higher resolution of local tomography, we can resolve that the SPMB is composed of three separate low velocity bodies, which correlate with the locations of three main volcanic centers (e.g., L1 for Antofalla, L2 for Cerro Galan, L3 for Cerro Blanco). These crustal low velocity zones were also observed by ambient noise correlation (Ward et al., 2017; Delph et al., 2017). Further, the geometry of the low velocity bodies beneath the volcanoes are different. The low velocity body beneath Cerro Galan (L2) reaches the greatest depth as it extends down to a depth of 50 km into the crust (Figure 5, profile P2) in accord with previous studies (Bianchi et al., 2013; Calixto et al., 2013; Heit et al., 2014; Liang et al., 2014; Delph et al., 2017). In addition to the crustal low velocity regions, our tomographic images illuminate a high velocity body (H2) at 90 km depth, located just to the west of the low velocity anomaly (L2) beneath Cerro Galan (Figure 5, P1 and P2), whose existence has been shown to be robust by a series of synthetic tests (section 3.2). With respect to with the depth to the Moho (Heit et al., 2014) and the subducting slab (Mulcahy et al., 2014), we observe that this H2 high velocity body sits above the subducted slab and is located beneath the Moho. Furthermore, the crust above the H2 anomaly seems to be thinned, which might indicate the detachment of the lithosphere.

The high velocity body resolved here correlates well with the high velocity body (called "DB" in Bianchi et al., 2013) observed by the teleseismic body and surface wave tomography (Bianchi et al., 2013; Calixto et al., 2013), which is also recovered as a low attenuation anomaly at the similar location (Liang et al., 2013). However, the high velocity body in our model (67.5ºW–68ºW, 90 km depth) is located shallower than and slightly west of the "DB" body, which is near 67ºW at a depth of >100 km where our tomography model loses resolution (Figures 6 and 7). Consequently, we speculate that "H2" and "DB" might represent the western and eastern parts of a same but much larger delaminated lithospheric block which caused the upwelling of fluids and melts above the delaminated block and triggered the volcanic eruptions (see interpretation scenario in Figure 8a). The asthenospheric upwelling increases the ambient mantle temperature and, thus, is capable to explain also the slab seismicity gap in this region.

A high velocity anomaly was observed at depths of 100–150 km beneath the backarc volcano Cerro Tuzgle in the northern Puna and was interpreted similarly as a delaminated lithospheric block as well (Schurr et al., 2006). In Figure S8 we plotted the two tomographic images at 105 km depth together. The high velocity anomaly closely beneath Cerro Tuzgle is deeper than and spatially separated from the H2 anomaly. Possibly, the Cerro Tuzgle volcanic center was triggered by a similar delamination process in the northern Puna above a subducting slab with a steeper subduction angle than in the southern Puna, the volume of which is probably much smaller than in the south, though.

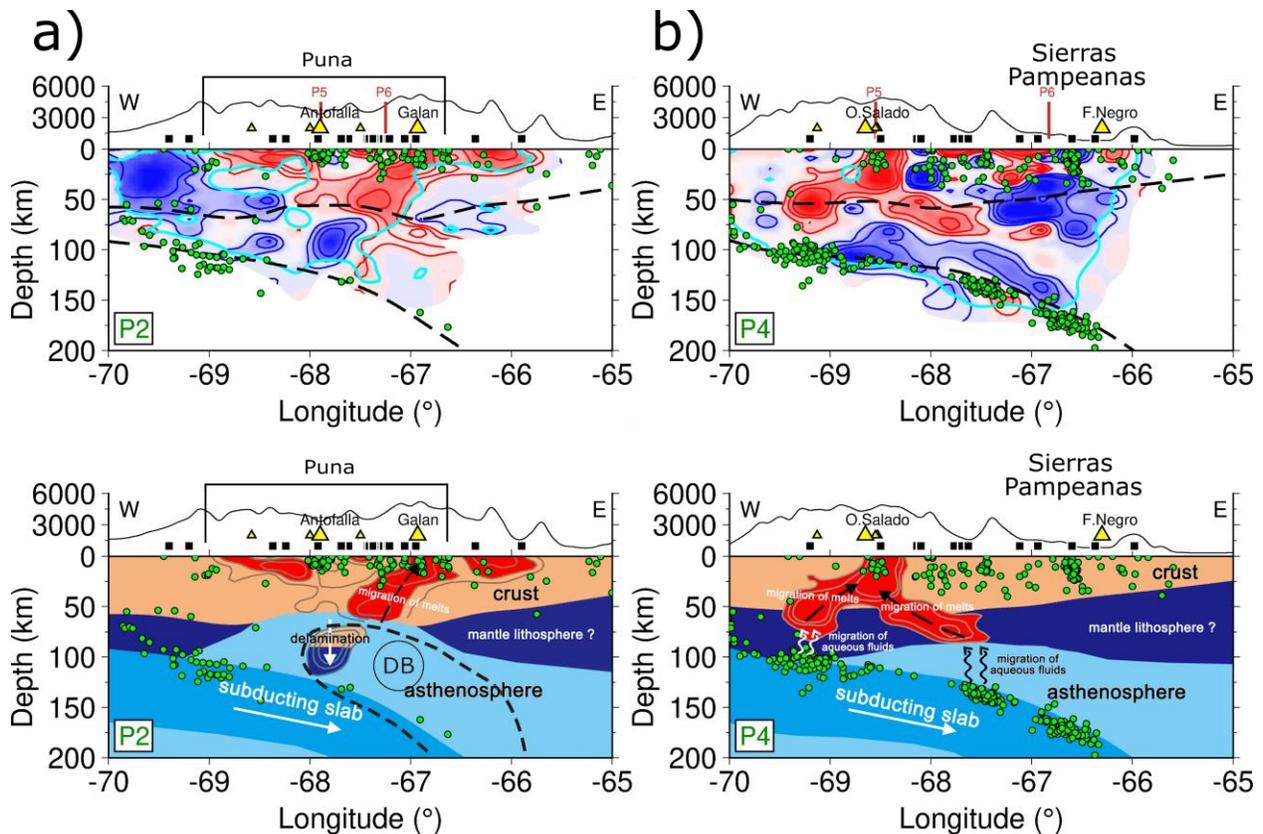

**Figure 8.** Vp perturbations and interpretation cartoons along cross sections P2 (a) and P4 (b). (a) The high velocity body "H2" is interpreted as the delaminated lithosphere, which caused the upwelling of fluids and melts into the crust and triggered the volcano eruption of Cerro Galan. The black dashed line outlines the high velocity anomaly from Calixto et al. (2013). The circle labeled DB is the high velocity anomaly from Bianchi et al. (2013). (b) Water was released from the slab near two earthquake clusters at 100 km (c1) and 150 km depth (c2), respectively, due to dehydration reaction. Fluid fluxes in the mantle wedge along two paths, which caused the decompression melting in the crust beneath the Ojos del Salado volcanic center.

## 4.2. Melt transitions at the southern termination of the plateau

In addition to the low velocity anomalies near the Galan volcanic center, which are thought to be related to the ongoing delamination, another interesting tectonic feature illuminated here is a geometrically complex low velocity anomaly beneath the region of the Central Volcanic Zone (Ojos del Salado volcanic center) in the frontal arc near the southern termination of the Puna plateau (Figure 4a, L4 and Figure 5a, profile P4). Here, two low velocity seismic branches at a depth of 50-80 km connect with a shallower low velocity seismic body beneath the Ojos del Salado volcanic center. In detail, the western branch extends towards a slab earthquake cluster at ~100 km depth that can be related to the current subduction (cluster c1 in profile P4). The other branch further east also extends into the uppermost mantle where it points to an earthquake cluster at a depth of ~150 km (labeled c2 in profile P4) that can also be related to the currently subducting slab. The synthetic test in Figure 7 shows the robustness of the inversion of these two low velocity branches. At the same places above the two slab earthquake clusters, Liang et al. (2014) observed high attenuation tomographic anomalies. In addition, a low velocity anomaly in a similar region beneath Ojos del Salado was detected by the joint teleseismic and regional P wave inversion analyses reported by Bianchi et al. (2013) although the shape of the anomaly was less well imaged due to resolution issues. Crustal low velocity anomalies are also observed beneath main volcanic centers by the ambient noise tomography (Ward et al., 2017; Delph et al., 2017). We performed a synthetic test to guarantee the robustness of the inversion for these two separated low velocity branches (Figure 7).

A model that has been discussed by Hacker et al. (2003) among others relates intermediate-depth intra-slab earthquakes to metamorphic dehydration reactions in the subducting slab. In this model, water is released from the oceanic slab and induces partial melting in the overlying mantle by lowering the solidus (Tatsumi, 1986). There are a number of studies that correlate well with this model. Based on the relationship between arc volcanism in the Alaska and New Zealand subduction zones and the 100 km depth contour of the associated Wadati-Benioff zones, Wiemer and Benoit (1996) found a positive anomaly of frequency-magnitude distribution (b-value) located at 90-100 km depth on the upper surface of the Wadati-Benioff zone. The authors suggested that this high anomaly of b-value was caused by the increase of pore pressure due to slab dehydration reactions. Increased pore pressure would lower the liquidus in the overlying asthenosphere and lead to the volcanism that occurs directly above this zone.

In contrast, Wyss et al. (2001) proposed a model that the magma was generated at 140-150 km depth and supplied for subduction zone volcanism in the northeastern Japan. Instead of rising straight up to the volcanoes from the nearest point of the deep seismic zone, the magma is transported along an inclined path from 150 km depth to the volcanoes. Lastly, in the north of our study region, attenuation tomography was performed to analyze the transportation of fluid and melt in the central Andean subduction zone (Schurr et al., 2003). At 24.2°S, a low Qp region is interpreted as a rough trace of melt ascent through the mantle wedge, whose sources are two distinct earthquake clusters at ~110 km and ~200 km depth. Additionally, two branches of this low Qp region indicate a complex transportation pattern, suggesting that ascent paths can be vertical upward but can also diverge back along the slab.

There are similarities between our images (Figure 5, P4) and those in the studies discussed above. They include: 1) two branches of low velocity anomalies from the same volcanic center towards two distinct earthquake clusters; 2) two intermediate-depth earthquake clusters (~100 km and ~150 km) most likely linked with the dehydration reactions of subducting Nazca plate; 3) a complex transportation pattern including a western straight upward path and eastern inclined path back along the slab. Given these observations, it seems reasonable to link the earthquakes clusters (c1 and c2) associated with the subducted slab to the volcanism beneath the Ojos del Salado volcanic center (see interpretation sketch in Figure 8b). In such an interpretation, the two low velocity branches (b1 and b2) may represent paths to transport melts and magma which are generated from the dehydration reactions in the slab. This transportation pattern seems to differ from that further north (Schurr et al., 2003) where melts and magma are interpreted to have transported to arc and backarc volcanism. By contrast, the two melt paths imaged here are concentrated beneath Ojos del Salado in the volcanic arc. However, as the Tuzgle volcanic center is geochemically similar to the Puna mafic volcanic backarc centers near Cerro Galan, it is more likely a product of asthenospheric upwelling and magma generation due to a lithospheric delamination process beneath the northern Puna, as evidenced by the high velocity anomaly beneath Cerro Tuzgle (Schurr et al., 2006).

## 5. Conclusions

By using local earthquake tomography, we have been able to image crustal and uppermost mantle anomalies and provide high quality images of the velocity distribution in the southern Puna plateau. Two prominent anomalies attract our attentions, providing reliable geophysical evidence for the delamination hypothesis suggested for our study region (Kay et al. 1994; Kay & Coira, 2009) and complex transportation patterns of melts in the southern Puna plateau:

The first is a high velocity body at around 90 km depth beneath the Moho and above the subducting slab underlying the region near the volcanic center of Galan. This high velocity body can be interpreted as a delaminated lithospheric block, which correlates well with the delamination evidence from previous studies (Bianchi et al., 2013; Calixto et al., 2013, 2014; Heit et al., 2014; Liang et al., 2014; Mulcahy et al., 2014). Due to high resolution obtained by using local earthquake tomography in this study, the shape and accurate position of this delaminated block is revealed. Together with a here resolved low velocity body in the crust near Cerro Galan, which likely represents a magma intrusion, it provides reliable evidence for the delamination hypothesis proposed by Kay et al. (2014) to explain the high topography of the southern Puna plateau with a thin crust and large volumes of ignimbrite deposits.

The second prominent velocity anomaly encountered is a geometrically complex low velocity body in the crust beneath volcanic center Ojos del Salado at the southern termination of the Puna plateau. The shape and the position of this low velocity anomaly are consistent with the high attenuation anomalies close to two intermediate-depth earthquake clusters of the slab (Liang et al., 2014) and suggests a connection between them. We propose that water is released due to slab dehydration reactions at 100 to 150 km depth, which would lower the solidus in the overlying asthenosphere. Hot mantle material migrates along a vertical and an inclined path, imaged here as two separate low velocity anomalies, respectively, to the crust and experience decompression melting while it rises.

## Acknowledgments

This work was supported by the Deutsche Forschungsgemeinschaft, the National Key R&D Program on Monitoring, Early Warning and Prevention of Major Natural Disaster (Grant No. 2017YFC1500301), the National Natural Science Foundation of China (Grant Nos. 11871297, U1839206), Tsinghua University Initiative Scientific Research Program and China Scholarship Council. The equipment for the experiment has been provided by the Geophysical Instrument Pool Potsdam (GIPP), IRIS PASSCAL and the Universities of Missouri and St Louis. Waveform data are archived at the GEOFON and IRIS data centers.

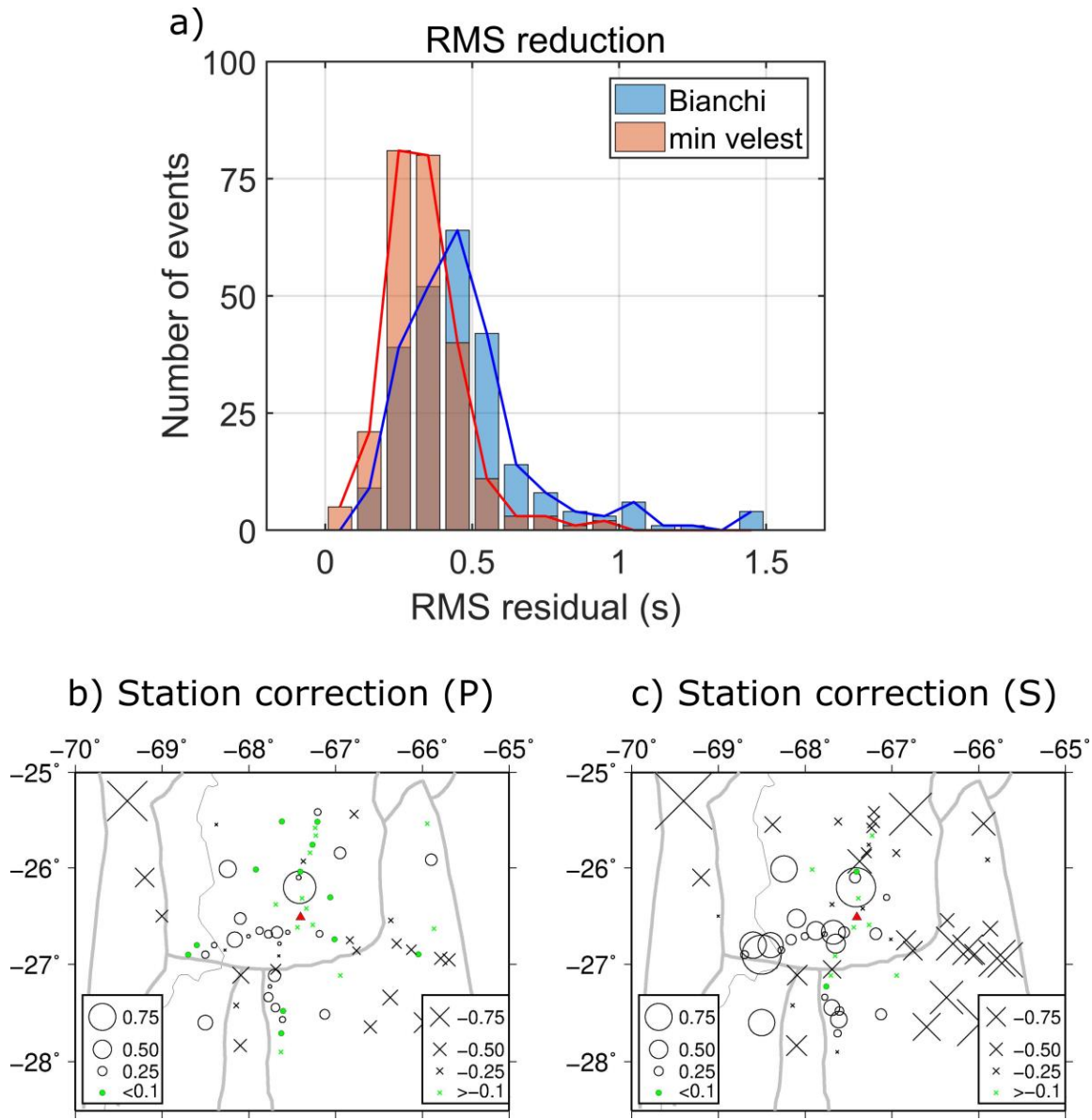

**Figure S1.** (a) RMS residuals of earthquakes located based on the Bianchi's model (blue) and the minimum 1D velocity model (orange). (b) and (c) Station corrections (in seconds) for P-picks and S-picks, relative to the reference station NS05 (red triangle). Positive values are indicated by circles, negative values are indicated by crosses.

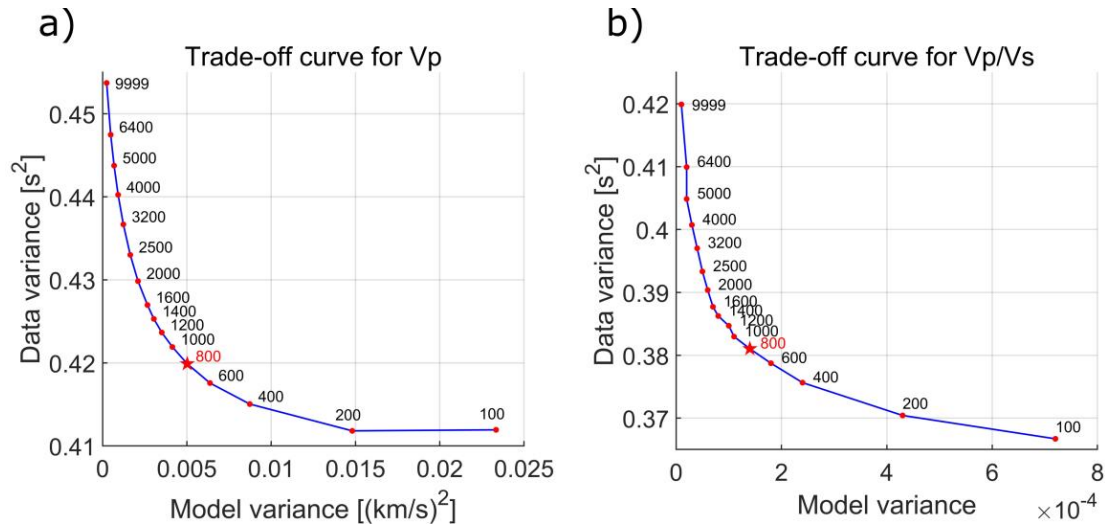

**Figure S2.** Trade-off curves of data variance vs. model variance for the Vp (a) and Vp/Vs (b) model, respectively. Suitable damping values (red pentagrams) are determined to make a balance between the model and data variance.

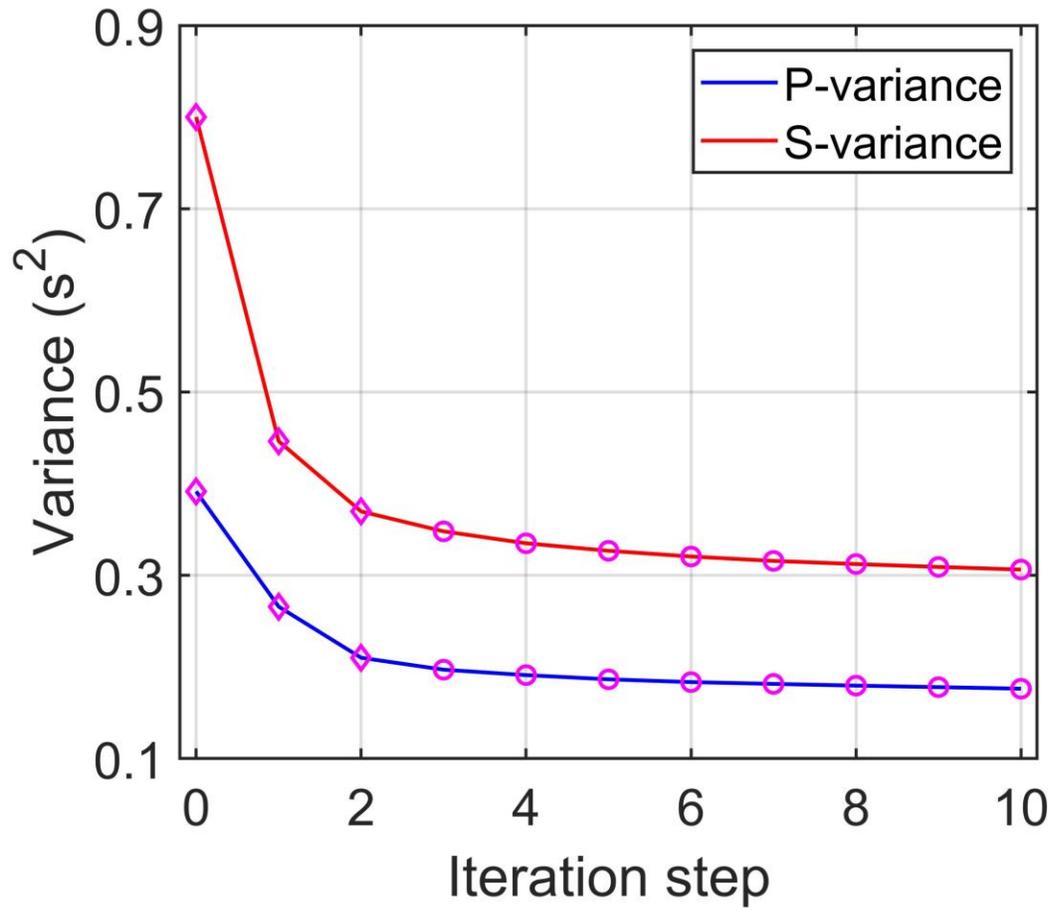

**Figure S3.** Data variance reduction curves of P- and S-variance with respect to the iteration step.

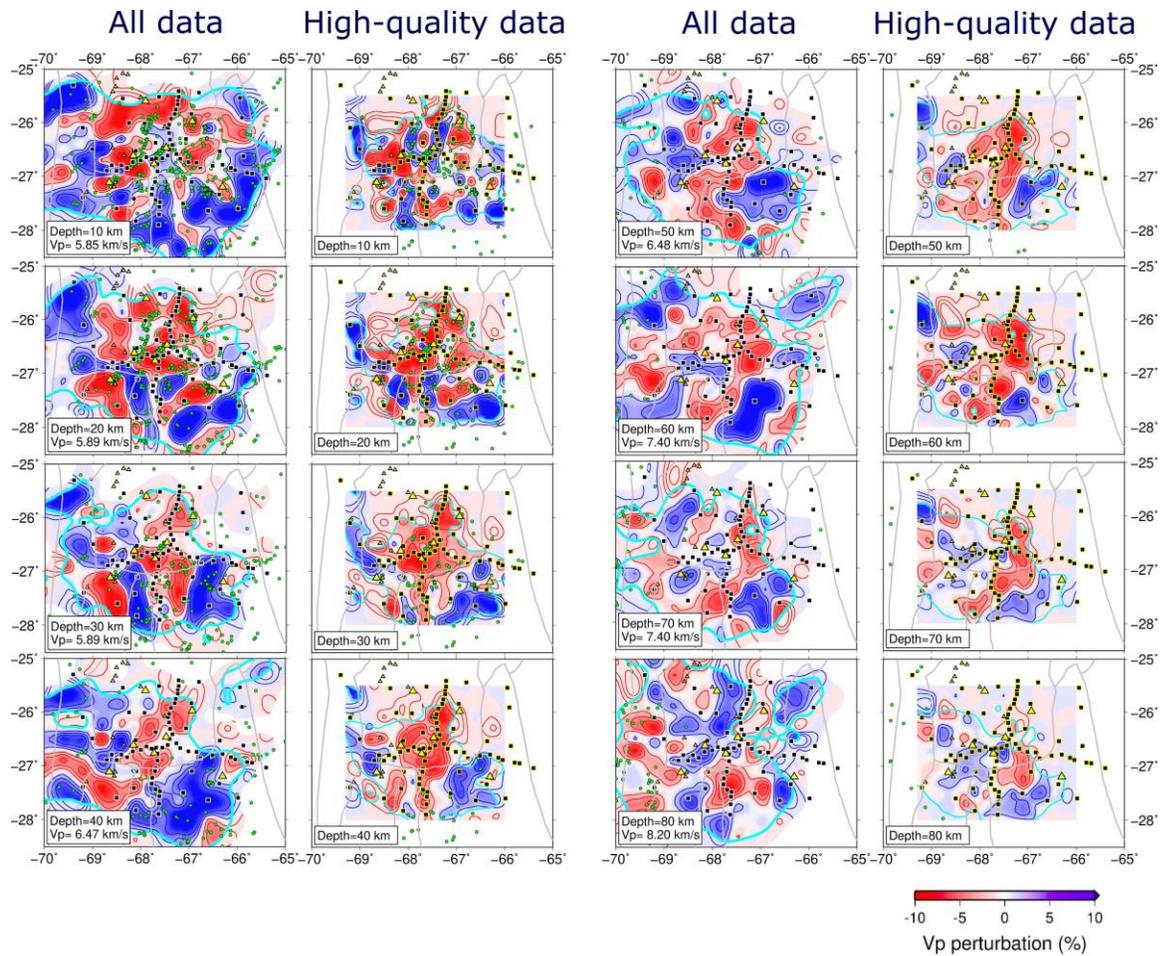

**Figure S4.** Comparison between our final result (Figure 4) and the inversion with only the 588 high-quality events involved in the 1D Vp inversion, whose azimuth gap is less than 240° and recorded by more than 10 stations. Due to a smaller number of events, the latter model has only resolution in a smaller region. In the central part of the study region above 40 km depth, where both models have resolution, the two velocity models are similar (e.g., the crustal low velocity zone in the southern Puna plateau and the high velocity anomaly beneath Sierras Pampeanas), indicating the robustness of the inversion of Figure 4 albeit events with larger azimuthal gap are used. However, due to the poor ray coverage, the delaminated body at 80 km can not be recovered in the latter model.

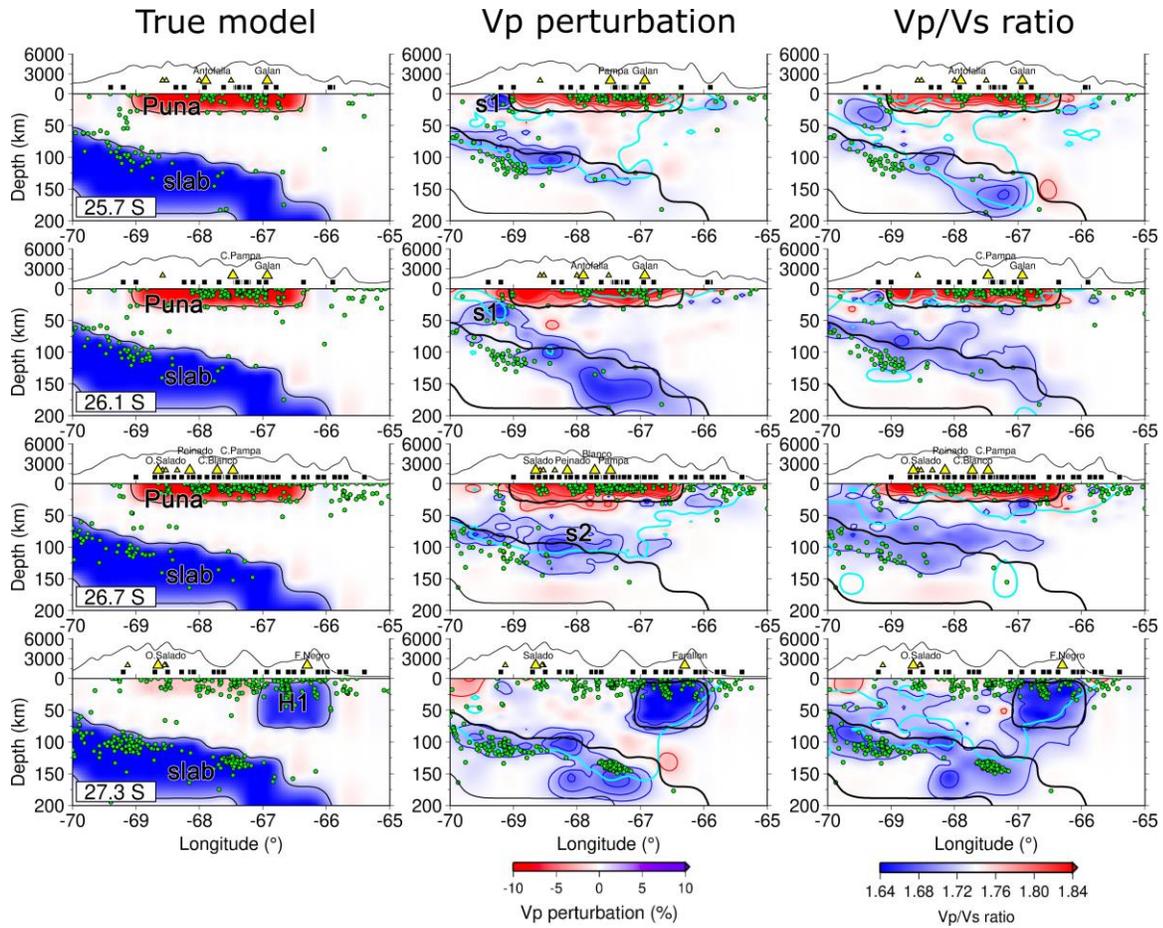

**Figure S5.** Same resolution test as shown in Figure 7 but excluding the two anomalies of "H2" and "L4". These removed anomalies are not recovered in the inversion results, indicating the robustness of these two velocity anomalies.

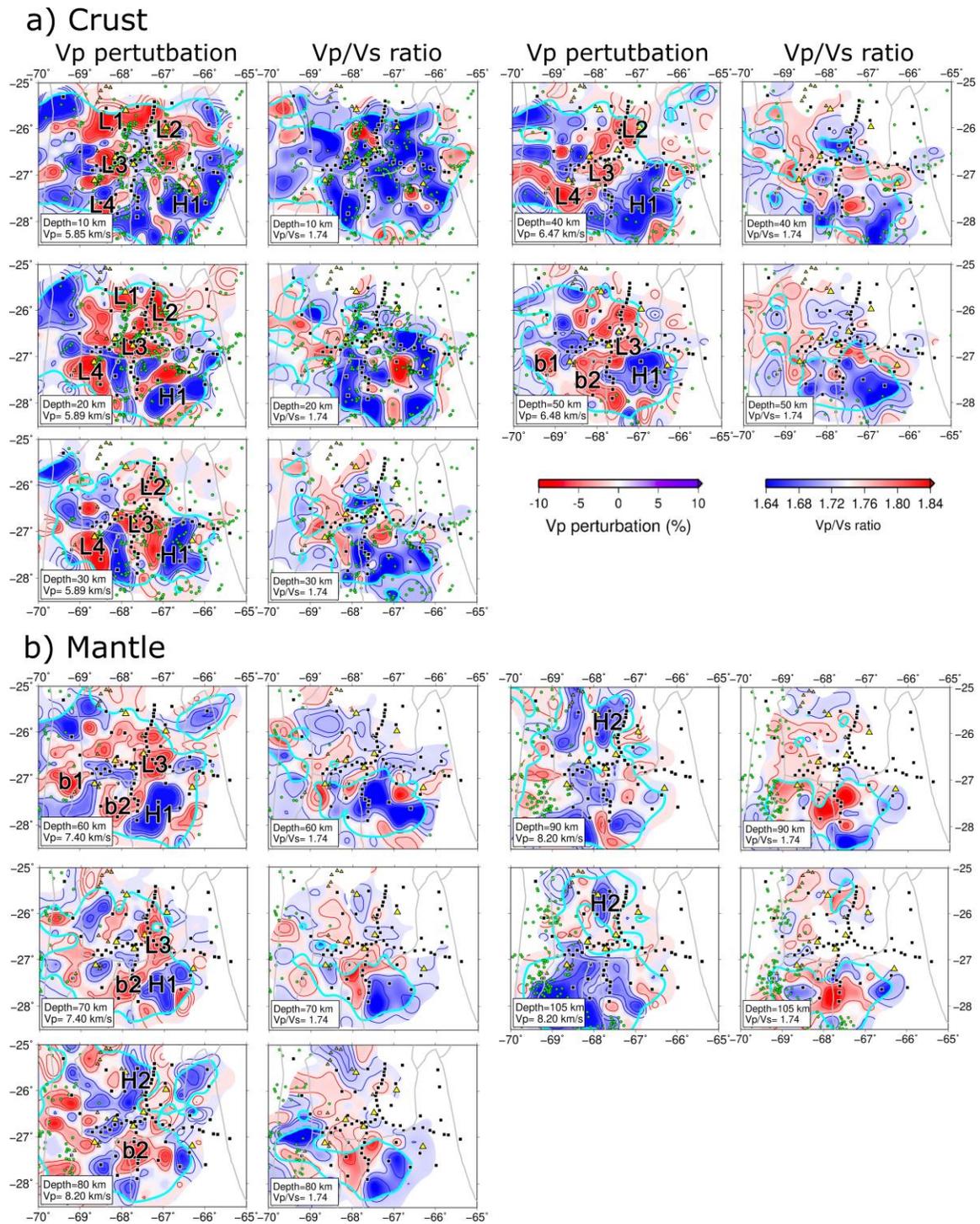

**Figure S6.** Horizontal sections of inversion results including Vp perturbation and Vp/Vs ratio models. All labels and features are the same as in Figure 4.

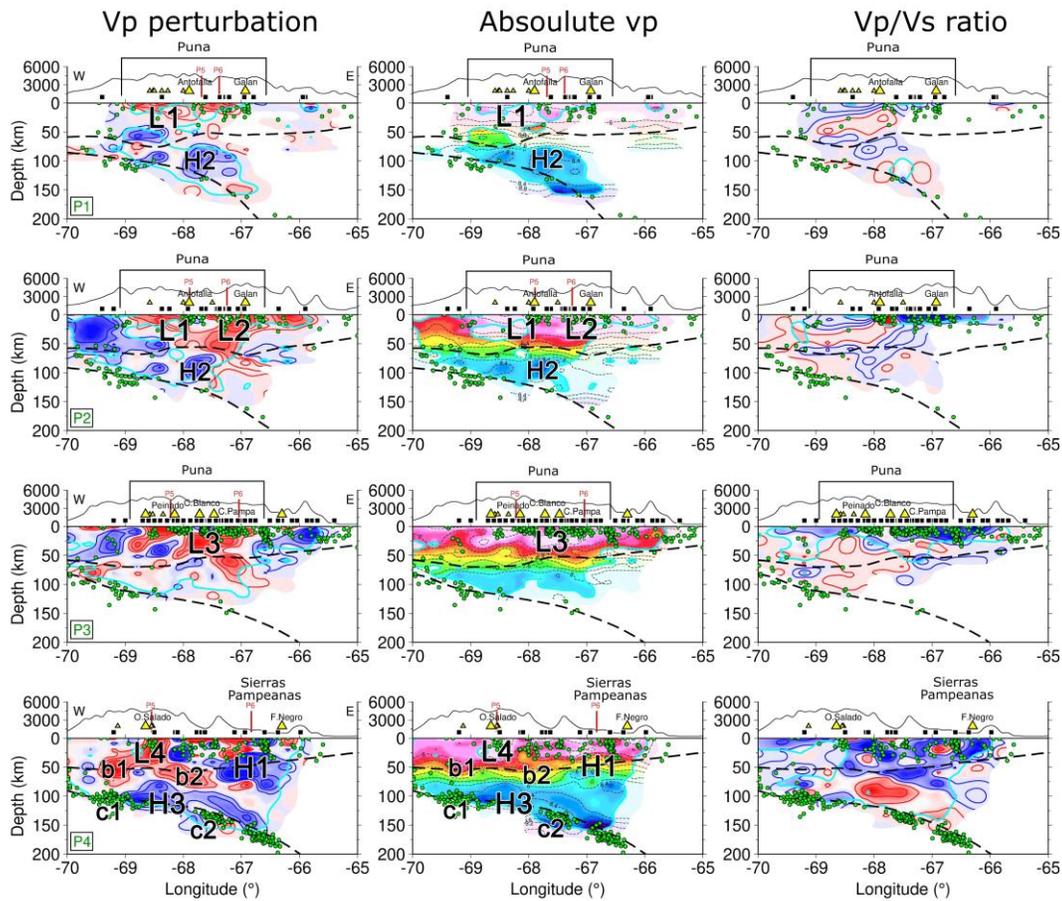

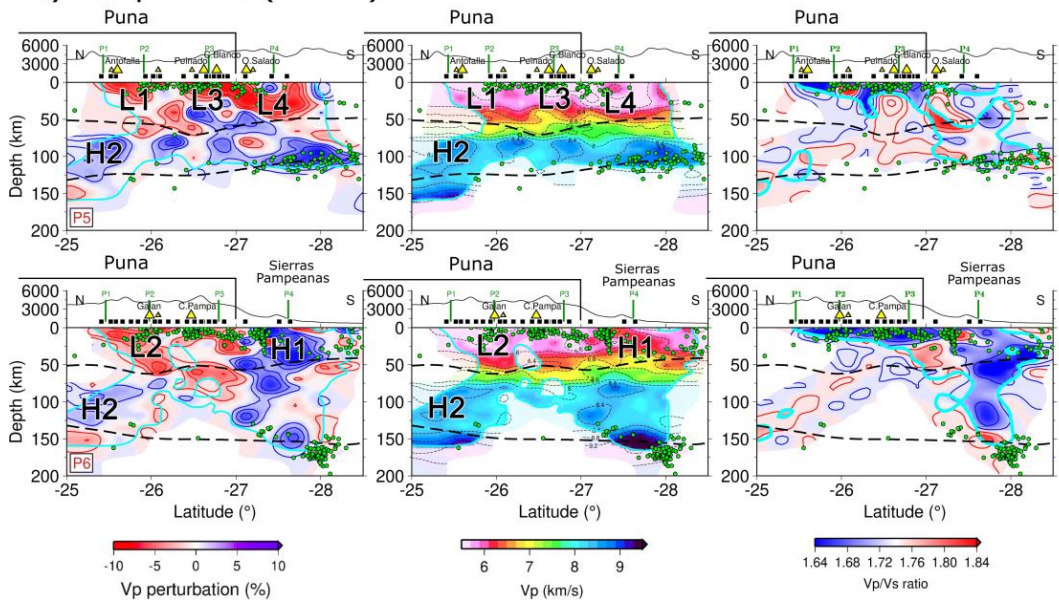

**Figure S7.** Vertical sections of inversion results including Vp perturbation, absolute Vp and Vp/Vs ratio models. All labels and features are the same as in Figure 5.

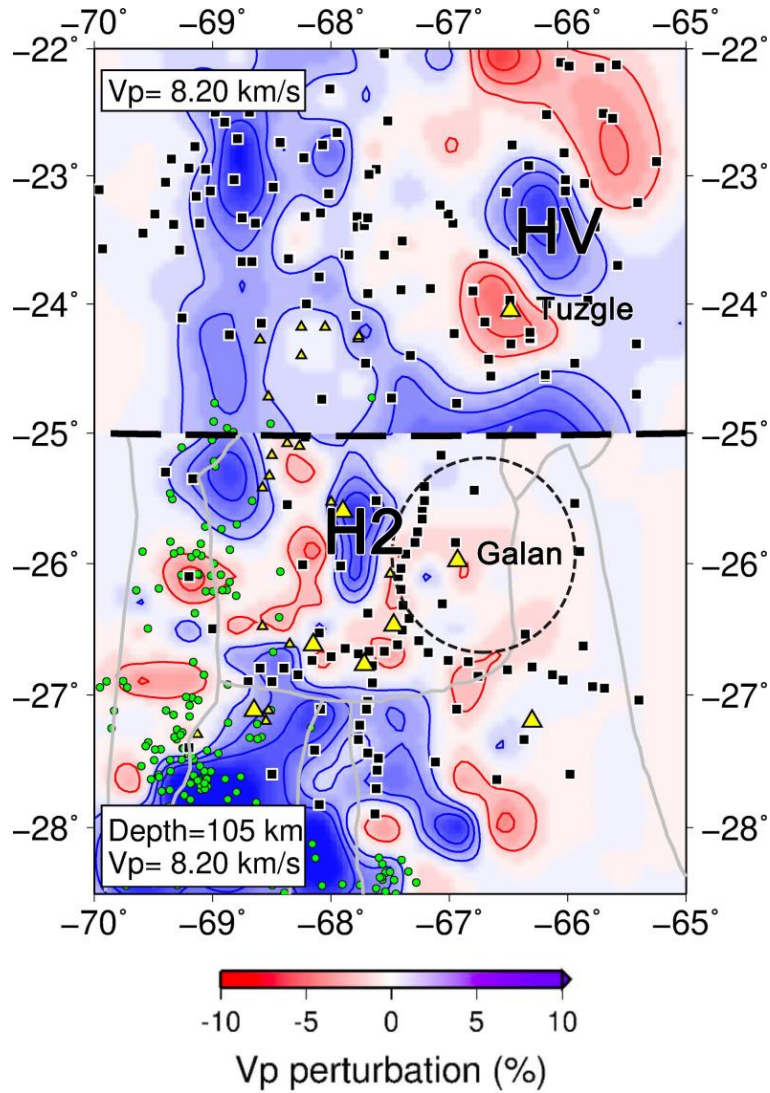

**Figure S8.** Combination of our inversion results with the tomography results in the northern Puna obtained by Schurr et al. (2006). The W-E black dashed lines at 25°S mark the boundary between the two models. The delaminated block detected in the northern Puna (HV) is closed related to the Cerro Tuzgle backarc volcano, which is separated from the delaminated block near Cerro Galan in the southern Puna (H2). The the black dashed circle indicates the delaminated block observed by Calixto et al. (2013).